\documentclass[superscriptaddress,twocolumn,prl,showpacs,floatfix]{revtex4}

\usepackage{dcolumn}
\usepackage{amsfonts}
\usepackage{amsmath}
\usepackage{amssymb}
\usepackage{bm}
\usepackage[dvips]{graphicx}
\usepackage[usenames]{color}

\begin{document}

\newcommand{\brm}[1]{\bm{{\rm #1}}}
\newcommand{\xv}{\bm{{\rm x}}}
\newcommand{\Rv}{\bm{{\rm R}}}
\newcommand{\uv}{\bm{{\rm u}}}
\newcommand{\nv}{\bm{{\rm n}}}
\newcommand{\ev}{\bm{{\rm e}}}

\title{Lipid Segregation on Cylindrically and Spherically Curved Membranes}

\author{Fangfu Ye}
\affiliation{School of Physics, Georgia Institute of Technology, Atlanta, GA 30332, USA}
\affiliation{Liquid Crystal Institute, Kent State University, Kent, OH 44242, USA}

\author{Robin L. B. Selinger}
\affiliation{Liquid Crystal Institute, Kent State University, Kent, OH 44242, USA}

\author{Jonathan V. Selinger}
\affiliation{Liquid Crystal Institute, Kent State University, Kent, OH 44242, USA}

\vspace{10mm}
\date{March 13, 2013}

\begin{abstract}
We investigate how an externally imposed curvature influences lipid segregation on two-phase-coexistent membranes. We show that the bending-modulus contrast of the two phases and the curvature act together to yield a reduced effective line tension. On largely curved membranes, a state of multiple domains (or rafts) forms due to a mechanism analogous to that causing magnetic-vortex formation in type-II superconductors. We determine the criterion for such multi-domain state to occur; we then  calculate respectively the size of the domains formed on cylindrically and spherically curved membranes.
\end{abstract}


\maketitle

\emph{Introduction}\ \ \ \
Lipid membranes play important roles in maintaining cell integrity and intracellular trafficking \cite{vanMeer}. At high temperatures, the three main components of the membranes (saturated lipids, unsaturated phospholipids and cholesterols) form a homogeneous mixture; below a critical demixing
temperature, the three components segregate into two coexistent fluid phases, a saturated-lipid-enriched liquid-ordered ($L_{o}$) phase and unsaturated-phospholipid-enriched liquid-disordered ($L_{d}$) phase. Small $L_o$ domains formed on cell membranes are also referred to as rafts \cite{simons}. The $L_{o}$ phase (or rafts) has a larger bending modulus than the $L_{d}$ phase \cite{veatch, baumgart, baumgart2}. There have been a large body of studies \cite{tae-young, groves, lipowsky, powers, ma, sorre, heinrich, stogbauer, longo, muller, jiang} investigating
how membrane curvature influences the segregation of membrane lipids. All these studies
are based on a physical mechanism that the $L_o$ phase because of its larger resistance to bending prefers flatter regions and drives
the $L_d$ phase to more curved regions,
i.e., these studies have implicitly assumed what matters is the variation of curvature rather than the curvature itself.

In this letter, we investigate how a uniform externally imposed curvature may influence lipid segregation on an $L_o$-$L_d$ coexistent membrane. We show that, with the presence of a bending-modulus contrast, the curvature of the membrane induces the lipids to tilt away from the membrane normal and yields a reduced effective line tension between the lipid domains; when the curvature is larger than a certain critical value, lipid segregation leads to the formation of multiple $L_o$/$L_d$ domains (or rafts) of microscopic lengthscale rather than a complete separation of the two phases. We determine the criterion for the multi-domain pattern to occur; we then calculate respectively the size of the domains formed on cylindrically curved and spherically curved lipid membranes. We point out that although inspired by lipid systems the results obtained in this letter apply to any membranes that contain two distinct smectic phases.
\\

\begin{figure}
	\centering
      \includegraphics[width=0.35\textwidth]{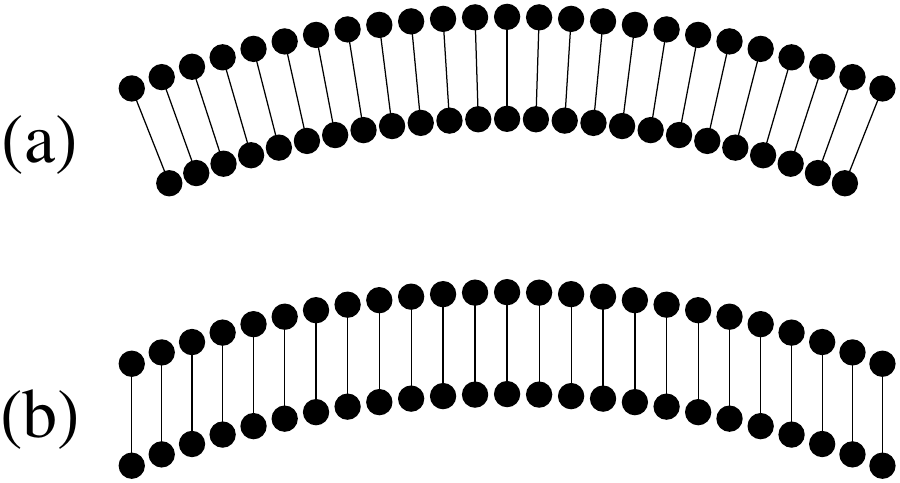}		
	\caption{
	Two types of lipid configurations of curved membranes. The dimers represent lipids, either lipids in bilayer-membranes or lipids in unilamellar systems. In (a), the lipids are perpendicular to the membrane surface; in (b), the lipids are parallel to each other.}
	\label{fig:bend}
\end{figure}

\emph{General Picture}\ \ \ \    Nonzero curvatures of lipid membranes can be either induced by their intrinsic spontaneous curvatures or imposed externally.
We consider the latter case\cite{incurv}. Examples of lipid membranes with an externally imposed curvature includes \emph{in vivo} membranes attached to BAR domains of proteins \cite{mcmahon} and \emph{in vitro} membranes attached to rigid substrates.
Fig.~\ref{fig:bend} shows two extreme types of lipid configurations resulting from an externally imposed curvature.
In Fig.~\ref{fig:bend}(a), the lipids are perpendicular to the membrane surface and the energy cost originates from the splay energy cost of lipid orientation; in Fig.~\ref{fig:bend}(b), the lipids are parallel to each other, and the energy penalty is induced by the deviation of lipids' orientations from the membrane normals, or more microscopically, by the relative sliding of the lipids with respect to one another.

In general, the lipid configuration of a curved membrane should be a superposition of the aforementioned two types. The corresponding elastic energy density thus includes two parts, the splay energy and the tilt energy penalty,
\begin{equation}\label{eq:f}
f=\tfrac{1}{2}K({\nabla'}\cdot\nv)^2-\tfrac{1}{2}C(\nv\cdot\uv)^2,
\end{equation}
where $\nv$ and $\uv$ are unit vectors representing, respectively, lipid orientation and membrane normal, $\nabla'$ represents the  two-dimensional derivative on a curved surface\cite{OuYangBook}, $K$ is the splay coefficient, and $C$ is a coefficient measuring how strongly $\nv$ is locked to $\uv$.
The combination of $K$ and $C$ gives a characteristic length $\xi=\sqrt{K/C}$, which is the penetration length of lipid tilting. For a single-phase membrane, given that $\xi$ is usually much smaller than the membrane size, most of the lipids are perpendicular to the membrane surface [see Fig.~\ref{fig:bend}(a)], and the bending modulus of the membrane is mostly determined by the splay coefficient $K$. A bending-modulus contrast in literature thus corresponds to a splay-coefficient contrast. Note that we have ignored in Eq.~(\ref{eq:f}) a divergence term because including it would not change our results qualitatively.

For an $L_o$-$L_d$-coexistent membrane, the total energy includes the contributions from both phases. We will use subscripts $o$ and $d$ to distinguish the quantities of these two phases. For convenience in the expressions, we will also use the subscript $i$, with $i=1$ representing the phase with smaller volume fraction and $i=2$ representing the one with larger volume fraction; in a case of no ambiguity, we will use a symbol without subscript to represent the corresponding quantities of both phases.
The total energy of a two-phase-coexistent membrane should also include a phase-boundary energy cost, which equals the product of the line tension $t$ and the phase-boundary length.
To minimize the boundary energy, the two phases on a flat membrane completely separate from each other, with each phase forming a large single domain. However, for a curved membrane, such a single domain may become unstable and split into multiple domains.

\begin{figure}
	\centering
      \includegraphics[width=0.35\textwidth]{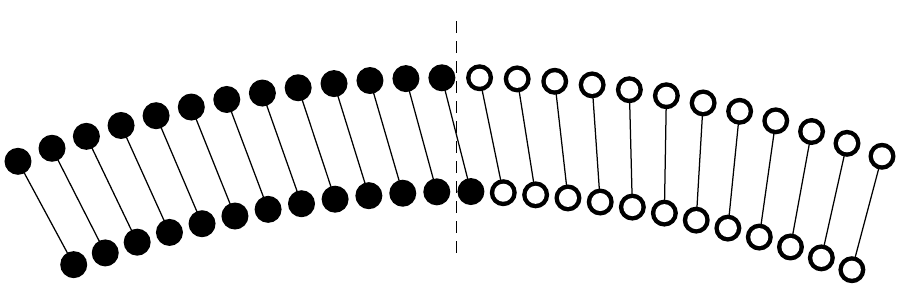}		
	\caption{
	Illustration of external-curvature-induced lipid-tilting in the phase-boundary region. The solid and hollow dimers represent, respectively, $L_o$ and $L_d$ lipids.  The dashed line represents the membrane normal at the phase boundary.}
	\label{fig:tilt}
\end{figure}

We now present the reason for the splitting of large single domains and determine the corresponding instability criterion. As illustrated in Fig.~\ref{fig:tilt}, when subjected to an externally imposed curvature, the $L_o$ lipids tilt away from the membrane normals and become more parallel to each other so as to reduce the splay energy cost, given that
$K_o$ is larger than $K_d$. The anchoring energy, i.e., the second term in Eq.~(\ref{eq:f}), constraints the tilting to occur within a narrow region of size $\sim\xi$ around the phase boundary. The energy change induced by such tilting can be computed. To simplify calculation, we assume that the size of the large single domains resulting from a complete phase separation is much larger than $\xi$
and that the imposed curvature is much smaller than $1/\xi$. Thus, in the aforementioned phase-boundary region, the splay
$\nabla'\cdot\nv_i$ can be approximated by $2H+(-)^i\partial \gamma_i/\partial s_i$, where $H$ is the mean curvature, $s_i$ is the distance between the phase boundary and the lipid location in phase $i$ (i.e., the arc length of the geodesic perpendicular to the phase-boundary line), and $\gamma_i$ is the angle
between the membrane normal and the orientation of lipids of phase $i$. Note that the lipids now tilt in the plane defined by the membrane normal and the tangential vectors of these geodesics. The factor $(-)^i$ is added into the expression of $\nabla'\cdot\nv_i$ because of the opposite directionality of $s_1$ and $s_2$. Substituting the approximate expression of $\nabla'\cdot\nv_i$ into Eq.~(\ref{eq:f}) and taking the functional derivative yields $\gamma_i=\gamma_b \exp(-s_i/\xi_i)$, where $\gamma_b$ represents the deviation angle at the phase boundary.
The energy change resulting from the lipid tilting can then be obtained by minimizing over $\gamma_b$ the sum of the energies of $L_o$ and $L_d$. The combination of this energy change and the line tension $t$ produces a reduced effective line tension:
\begin{equation}\label{eq:te}
t_e=t-\frac{2H^2(K_o-K_d)^2}{\sqrt{C_oK_o}+\sqrt{C_dK_d}}.
\end{equation}
Eq.~(\ref{eq:te}) clearly shows $t_e$ can be lowered either by increasing $H$ or the contrast between $K_o$ and $K_d$ or by decreasing $C_o$ and $C_d$.

The instability criterion of the single-domain state can then accordingly be obtained by simply setting $t_e=0$. We can further simplify this criterion to an approximate form: $t/H^2\alt K\xi$.
Given that $K$ is of order $10^{-19}\textrm{J}$ for most lipid systems \cite{OuYangBook}
and $\xi$ can be assumed to be comparable to the membrane thickness and is thus of order $1\textrm{nm}$ \cite{Kozlov},
the criterion is then $t/H^2\alt10^{-28}\textrm{J}\cdot\textrm{m}$. Depending on the compositions and temperature, the line tension $t$ on a lipid membrane usually varies between $1\textrm{pN}$ and $0.01\textrm{pN}$ (or even smaller) \cite{baumgart, baumgart2, Benvegnu, Schick}.
For $t\sim1\textrm{pN}$, the critical curvature $H_c$ is thus of order $10^8\textrm{m}^{-1}$; for $t\sim0.01\textrm{pN}$ (for example, in systems with presence of hybrid lipids \cite{Brewster}), $H_c$ is of order $10^{7}\textrm{m}^{-1}$.

As $t_e$ becomes negative, the large single domains resulting from a complete phase separation become unstable and split into multiple small domains.
In the following sections,
we calculate the size of the lipid domains formed on cylindrically curved and spherically curved membranes, respectively.
\\

\begin{figure}
	\centering
      \includegraphics[width=0.3\textwidth, height=0.25\textwidth]{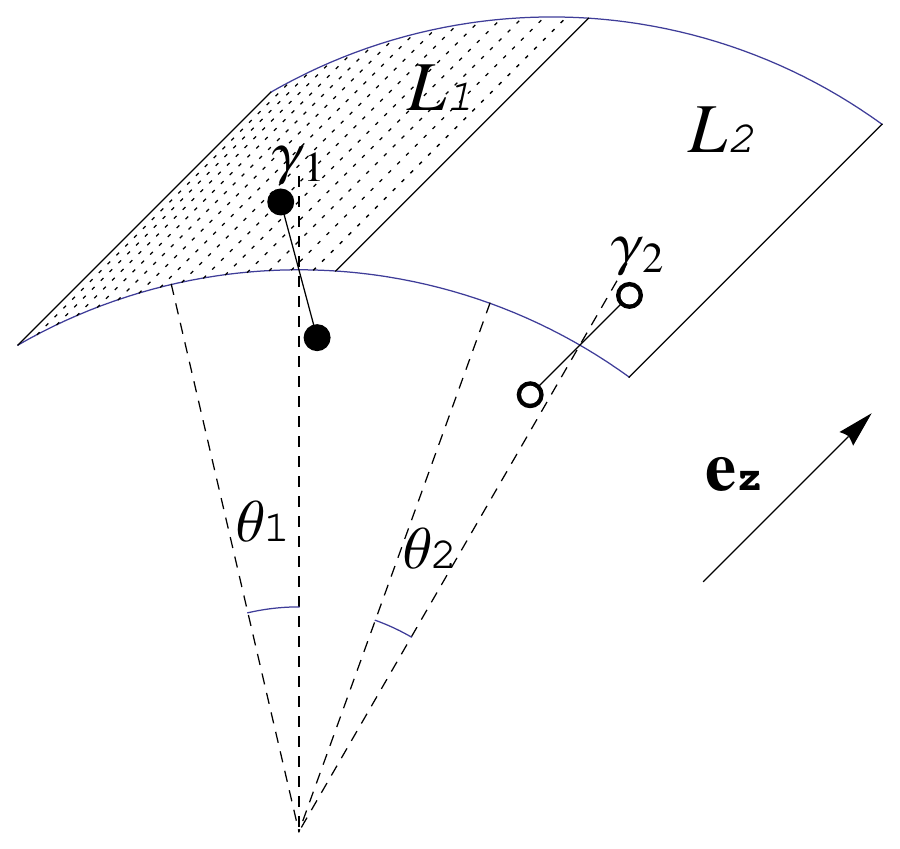}		
	\caption{
	Cartoon of $L_1$ and $L_2$ domains on a cylindrical surface.}
\label{fig:ccartoon}
\end{figure}

\emph{Lipid Segregation on a Cylindrically Curved Membrane}\ \ \ \
For a lipid membrane subjected to a uniform cylindrical curvature, we assume there is translational symmetry along $\ev_z$, the direction of the long axis of the cylinder, and that the lipids align perpendicular to $\ev_z$. In this case, a domain is a strip along $\ev_z$, and the three-dimensional problem becomes a two-dimensional problem. The splay of the lipid orientation can then be expressed as
$\nabla'\cdot\nv_i=\cos\gamma_i[1+\partial\gamma_i/\partial\theta_i]/R_c$, where
$R_c$ is the radius of the cylinder and the $\theta$'s are the azimuthal angles representing the lipid location in a domain (see Fig.~\ref{fig:ccartoon}). In the center of a domain, i.e., at $\theta_i=0$, the lipids align along the membrane normal with $\gamma_i=0$. Substituting the expression of $\nabla'\cdot\nv_i$ into Eq.~(\ref{eq:f}) and then differentiating the energy density with respect to $\gamma_i$ yields
$\gamma_i=A_i\sinh{(\theta_i R_c/\xi_i)}$, where the $A$'s are the amplitudes and the approximations $\xi_i\ll R_c$ and $\gamma_i\ll 1$ have been used. Approximately speaking, the deviation $\gamma_i$ decays exponentially for domains of size much larger than $\xi_i$ and linearly for those of size smaller than $\xi_i$. The amplitudes $A_1$ and $A_2$ are not independent from each other---they are related by the continuity condition
$\gamma_1(\theta_1^s)=\gamma_2(-\theta_2^s)$,
where $\theta_1^s$ and $\theta_2^s$ are, respectively, the angular sizes of the $L_1$ and $L_2$ domains (from the domain centers to the domain edges).
We then minimize the total energy (per unit length along the long axis of the cylinder) over the $A$'s and obtain
\begin{equation}\label{eq:fcy2}
E=\frac{2\pi}{\theta^s_1+\theta^s_2}\Big[t-\frac{(K_1-K_2)^2/R_c^2}{2\sum_{i=1}^2\sqrt{C_iK_i}\coth(\theta^s_iR_c/\xi_i)}\Big],
\end{equation}
which clearly shows a competition between the energy penalty of the line tension and the energy gain of the director tilting induced by the difference between $K_1$ and $K_2$. The critical line tension $t_c$, below which the completely separated state becomes unstable, is given by the maximum value of the second term in the parenthesis in Eq.~(\ref{eq:fcy2}), which is the same as the one we can obtain from Eq.~(\ref{eq:te}) by setting $t_e=0$ and $H=1/(2R_c)$.

\begin{figure}
    \includegraphics[width=0.23\textwidth]{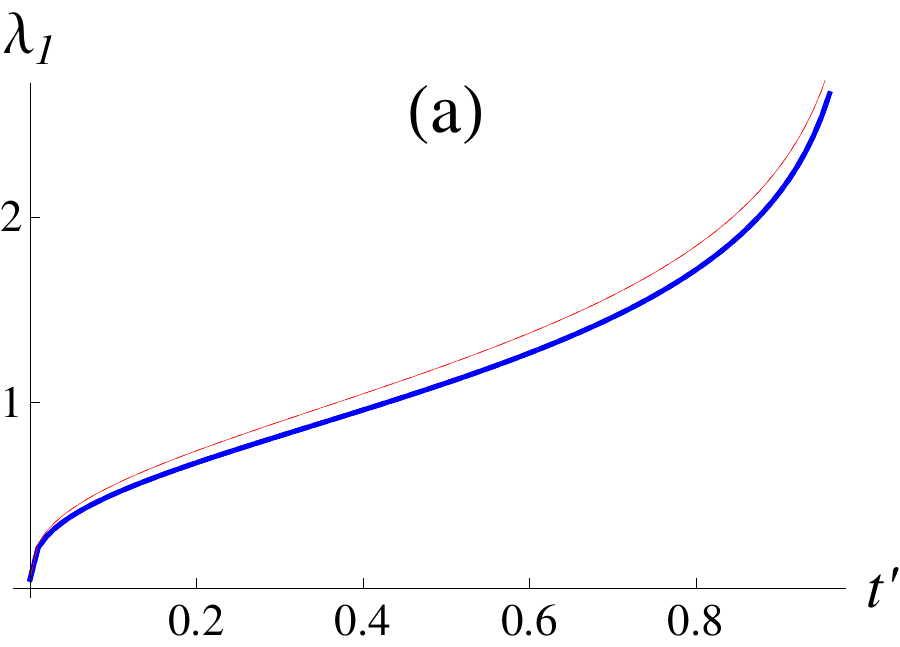}
    \hfill
    \includegraphics[width=0.23\textwidth]{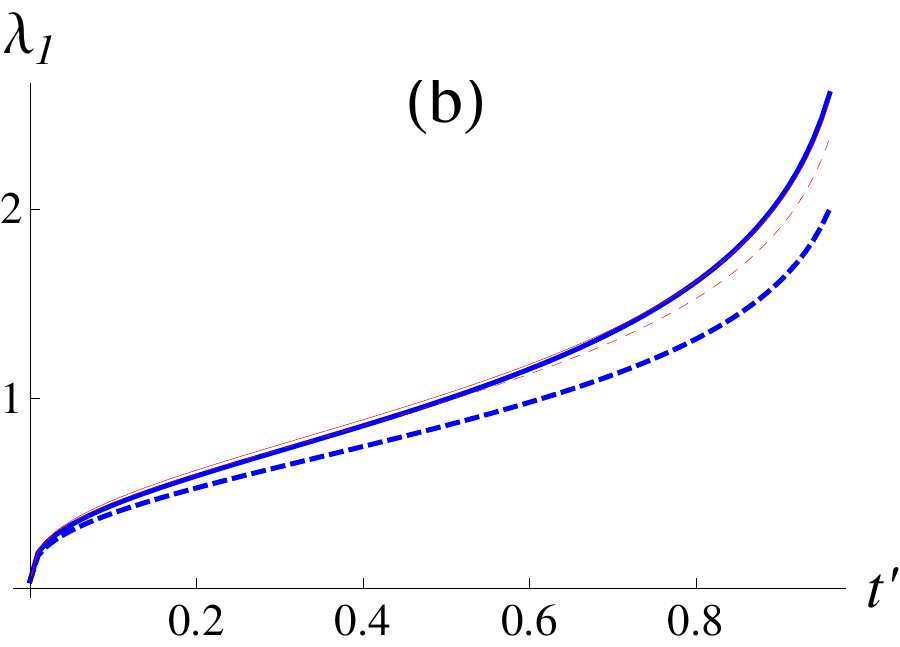}
    \includegraphics[width=0.23\textwidth]{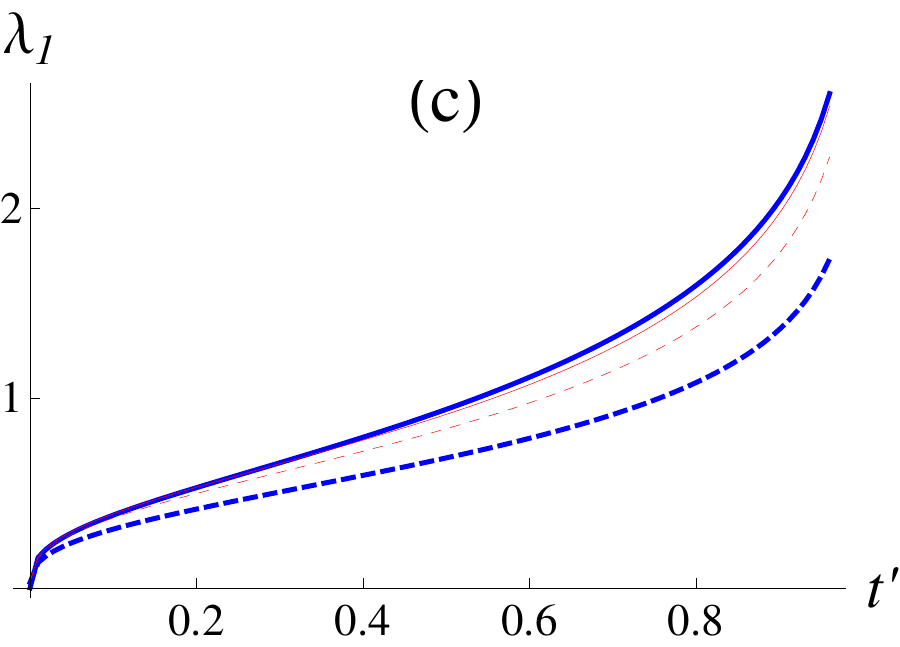}
    \hfill
    \includegraphics[width=0.23\textwidth]{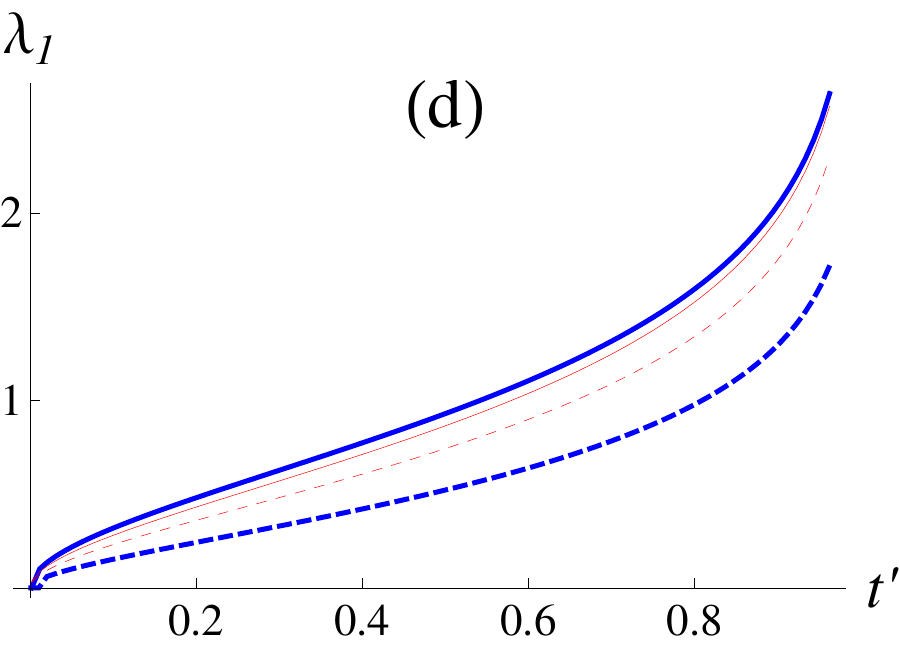}
	\caption{
	Relations between the $L_1$-domain size (measured in unit $\xi_o$, $\lambda_1=\theta_1^sR_c/\xi_o$) and the normalized line tension $t'=t/t_c$ on a cylindrically curved membrane for (a) $\phi=0.5$, (b) $\phi=0.4$, (c) $\phi=0.3$, and (d) $\phi=0.01$. The (blue) thick solid and dashed lines correspond to $K_1=2K_2$ and $K_1=K_2/2$, and the (red) thin solid and dashed lines correspond to $K_1=1.2K_2$ and $K_1=K_2/1.2$, respectively. We have set $C_1=C_2$.}
	\label{fig:csize}
\end{figure}

At $t<t_c$, a stable multi-domain ground state is expected. The domain size can be determined by numerically minimizing Eq.~(\ref{eq:fcy2}) over $\theta^s_1$ (or $\theta^s_2$) with the constraint $\theta^s_1/\theta^s_2=\phi/(1-\phi)$ applied, where $\phi$ is the volume fraction of phase~$1$.
The results are given in Fig.~\ref{fig:csize}, which shows that the domain size decreases with $t$ and quickly reduces to order $\xi_o$. For $\phi=1/2$ [see Fig.~\ref{fig:csize}(a)], because of the permutation symmetry of the two lipid phases, the curves describing how the domain size varies with the normalized line tension $t'$ ($=t/t_c$) for the $K$'s with a certain ratio $v$ (i.e., $K_1=vK_2$) overlap with the curves for the $K$'s with a ratio of $1/v$ (i.e., $K_1=K_2/v$). For $\phi<1/2$, this permutation symmetry is broken: the $L_o$ domains formed on membranes with the $L_o$ phase having a volume fraction $\phi$
have a larger size than the $L_d$ domains formed on the phase-permutated membranes, i.e., membranes with the $L_d$ phase having a volume fraction $\phi$.
Moreover, the size difference increases as $\phi$ decreases [see Fig.~\ref{fig:csize}(a)--(d)]. The origin of this size difference and its variation with $\phi$ is given as follows. As can be seen from Fig.~\ref{fig:tilt}, with the presence of an external curvature, the $L_o$ lipids provide tilt-driving forces while the $L_d$ lipids resist tilt. Membranes
with the $L_d$ phase having volume fraction $\phi$
possess more $L_o$ lipids and thus tend to have more phase-boundary regions so that tilt can occur. A preference for boundary regions means a preference for smaller domains, and therefore leads to the aforementioned domain-size difference and its variation with $\phi$.

In addition to the numerical results, we also present in the following analytical results for the small $t$ limit so as to clearly elucidate the dependence of the domain size (i.e., $\theta^s_1$ and $\theta^s_2$) on the parameters. We consider two cases separately: (i) the volume fractions of the two phases are of same order, i.e., $\phi\sim 1/2$; and (ii) one phase has a much smaller volume fraction than the other, i.e., $\phi\ll 1/2$.
In the case of $\phi\sim1/2$, when $t\ll t_c$, we have $\theta^s_i\ll\xi_i/R_c$ for both phases, and the deviation angles $\gamma_1$ and $\gamma_2$ both decay (approximately) linearly from the phase boundary to the domain centers. We thus approximate $\coth(x)$ in Eq.~(\ref{eq:fcy2}) as $1/x+x/3$. We then differentiate the energy given in Eq.~(\ref{eq:fcy2}) with respect to $\theta_1^s$, with the relation between $\theta_1^s$ and $\theta_2^s$ applied, and obtain
\begin{equation}\label{eq:theta1s1}
\theta_1^{s}\approx
\Big(\frac{3t}{R_c\tilde{C}}\Big)^{1/3}\Big[\frac{1+r\phi/(1-\phi)}{1-r}\Big]^{2/3},
\end{equation}
where $\tilde{C}= C_1+C_2(1-\phi)/{\phi}$ and $r= K_2/K_1$. We mention the following three noteworthy points contained in Eq.~(\ref{eq:theta1s1}):
(i) a large contrast between the $K$'s (i.e., large $|r-1|$) gives a small domain;
(ii) although the angular size $\theta^s_1$ decreases as $R_c$ increases, the linear domain size ($R_c\theta^s_1$) increases with $R_c$;
(iii) at $\phi=1/2$, the permutation symmetry is conserved, viz, the value of $\theta^s_1$ does not change under the replacement $r\rightarrow 1/r$.
We now turn to the case of $\phi\ll 1/2$. In this case, we have $\theta^s_1\ll \xi_1/R_c$ but $\theta^s_2\gg \xi_2/R_c$; $\gamma_1$ decays linearly and $\gamma_2$ decays exponentially from the phase boundary to the domain center. We thus replace $\coth(\theta_1^sR_c/\xi_1)$ and $\coth(\theta_2^sR_c/\xi_2)$ in Eq.~(\ref{eq:fcy2}) by $\xi_1/(\theta_1^sR_c)$ and $\coth(\infty)$, respectively. Minimizing Eq.~(\ref{eq:fcy2}) then yields
\begin{equation}\label{eq:theta1s2}
\theta_1^{s}\approx
\Big(\frac{2t}{\sqrt{C_2K_2}}\Big)^{1/2}\frac{1}{|1-r|},
\end{equation}
which is inversely proportional to $|r-1|$ and independent of the radius $R_c$ (the linear domain size is thus linearly proportional to $R_c$). In addition, the angular size is now more sensitive to the variation of $t$ than the angular size given in Eq.~(\ref{eq:theta1s1}) is.
\\

\emph{Lipid Segregation on a Spherically Curved Membrane}\ \ \ \
We proceed to address lipid segregation on a spherically curved membrane. A complete phase separation now leads to the formation of a single (curved) disk of $L_1$ lipids embedded in an $L_2$-lipid sea.
The instability criterion of the single-disk pattern an be obtained from Eq.~(\ref{eq:te}) by setting $t_e=0$ and $H=1/R_s$. For a negative $t_e$, a multi-domain pattern forms. In the case of $\phi\sim 1/2$, the shape of the formed multiple domains is complicated, and this complication makes an analytic calculation of the system energy inaccessible. We will study the $\phi\sim 1/2$ case
in a numerical approach in a future publication and consider here only
the case of $\phi\ll 1/2$. In this case, we have a pattern of multiple $L_1$ disks embedded in an $L_2$ sea.

We first give the explicit form of the energy as a function of the deviation angle $\gamma$
and the equation determining how $\gamma$ varies from the center of a domain to the domain edge.
We adopt a spherical-coordinate description for the positions of the lipids, with the point of zero polar angle (i.e., $\theta=0$)
corresponding to the center of an $L_1$ domain, and we assume $\gamma$ only depends on the polar angle $\theta$ but not the azimuth angle.
Thus,
for the total energy, we have
\begin{eqnarray}\label{eq:es}
E\!\!&=&\!\!2\pi N_d\Big\{R_s t \sin \theta_1^s
\nonumber\\
\!\!& &\!\!\
+R_s^2\int_0^{\theta_1^s}\!\!  d\theta \sin\theta \Big[\tfrac{1}{2}K_1(\nabla'\cdot\nv_1)^2\!
+\tfrac{1}{2}C_1 \sin^2 \gamma_1\Big]
\nonumber\\
\!\!& &\!\!\
+R_s^2\int_{\theta_1^s}^{\theta^u}\!\! d\theta \sin\theta \Big[\tfrac{1}{2}K_2(\nabla'\cdot\nv_2)^2\!
+\tfrac{1}{2}C_2 \sin^2 \gamma_2\Big] \Big\},\ \ \ \ \\
\nabla'\cdot\!\!\!&\nv_i&\!\!=
\frac{1}{R_s}\Big\{\cos\gamma_i(\theta)\Big[2+\frac{\partial\gamma_i(\theta)}{\partial\theta}\Big]
+\cot\theta\sin\gamma_i(\theta)\Big\},\nonumber
\end{eqnarray}
where $R_s$ is the radius of the sphere,
$\theta_1^s$ and $N_d=2\phi/(1-\cos \theta_1^s)$
are respectively the angular size and the total number of the $L_1$ domains. The upper limit $\theta^u$ of the second integration equals to $\pi$ in the single-domain case;
and its value in the multi-domain case will be addressed later.
Differentiating $E$ with respect to $\gamma$, under the assumption that $\gamma$ is small, yields
\begin{equation}\label{eq:fgamma}
\frac{\partial^2\gamma_i}{\partial \theta^2}
+\frac{\partial \gamma_i}{\partial\theta}\cot \theta
-\gamma_i \Big(\frac{R_s^2}{\xi_i^2}-4+{\csc^{2}\theta}\Big)=0.
\end{equation}

\begin{figure}
	\centering
\includegraphics[width=0.3\textwidth]{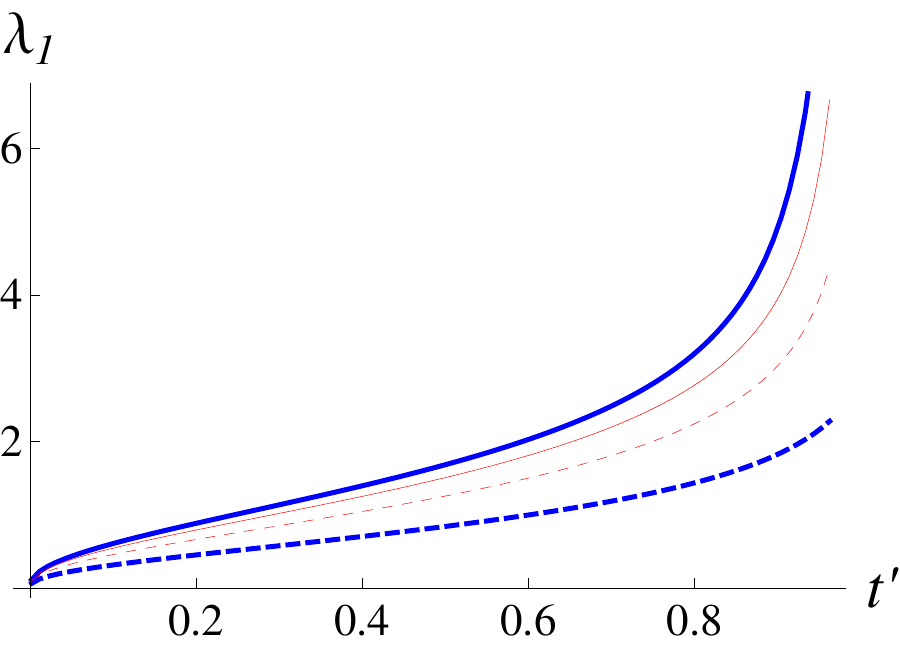}
	\caption{
	Relations between the $L_1$-domain size (measured in unit $\xi_o$, $\lambda_1=\theta_1^sR_s/\xi_o$) and the normalized line tension $t'=t/t_c$ on a spherically curved membrane for $K_1=2K_2$ [the (blue) thick solid line], $K_1=K_2/2$ [(blue) thick dashed], $K_1=1.2K_2$ [(red) thin solid] and $K_1=K_2/1.2$ [(red) thin dashed] with the assumptions $\phi\ll 1$ and $C_1=C_2$.}
\label{fig:ssize}
\end{figure}

We now proceed to determine the domain size, i.e., the size of the $L_1$ disks.
Given the constraint $\phi\ll 1/2$, the distance between the disks becomes much larger the disk size; and, in the $L_2$ sea, only a narrow annulus around the disks contributes to the system energy.
Furthermore, because the domain size is small, the deviation angle is nonzero only at small $\theta$'s.
We thus replace $\cot \theta$ and $\csc\theta$ in Eq.~(\ref{eq:fgamma}) by $1/\theta$, and obtain
$\gamma_1\varpropto{\rm BI}_1(b_1\theta)$ and $\gamma_2\varpropto{\rm BK}_1(b_2\theta)$,
where $b_i= \sqrt{(R_s/\xi_i)^2-4}\approx R_s/\xi_i$,
and ${\rm BI}_n(x)$ and ${\rm BK}_n(x)$ are, respectively, the modified Bessel functions of the first and second kind.
To calculate the energy, we then follow a procedure similar to that for the cylindrical case and obtain
\begin{subequations}
\begin{eqnarray}
& &\! E=
8\pi\phi\Big[\frac{R_st}{\theta_1^s}-\frac{2(K_1-K_2)^2}{{\rm W}(\theta_1^s)}\Big],\\
& &\! {\rm W}(\theta_1^s):=
K_1b_1\theta_1^s\frac{{\rm BI}_0(b_1\theta_1^s)}{{\rm BI}_1(b_1\theta_1^s)}
+K_2b_2\theta_1^s\frac{{\rm BK}_0(b_2\theta_1^s)}{{\rm BK}_1(b_2\theta_1^s)}.\ \ \ \ \ \ \ \
\end{eqnarray}
\end{subequations}
Minimizing this energy yields the domain size. Fig.~\ref{fig:ssize} shows the numerical results, which are qualitatively similar to the results of cylindrically curved membranes with small $\phi$ [see Fig.~\ref{fig:csize}(d)].
\\

\emph{Concluding Remarks}\ \ \ \
We have shown that a curved two-phase-coexistent lipid membrane, compared with a flat one, has a smaller effective line tension due to lipid tilting.
Moreover, a large enough curvature leads to a negative effective line tension, and thus induces a state with multiple domains (or rafts) of microscopic length scale determined by the splay and anchoring coefficients. We expect that the theoretical framework addressed in this paper, if applied to small intracellular vesicles, has some biological significance.

We mention that the multi-domain-inducing mechanism addressed in this paper is different from that using elastic repulsions to stabilize multiple domains on a lipid  vesicle \cite{baumgart, phillips}, but analogous to the mechanism via which a large magnetic field induces the formation of magnetic vortices in type-II superconductors \cite{Tinkham}. The tilt penetration length $\xi$ in this work is the counterpart of the magnetic field penetration length in superconductors. What corresponds to the superconducting coherence length is the correlation length of the density fluctuation of the various components of lipid membranes \cite{Schick}. We have implicitly assumed this correlation length is smaller than $\xi$, viz., we have assumed a sharp boundary between the two coexisting phases. We will present elsewhere how an externally imposed curvature influences lipid segregation on membranes with this assumption violated, for example, on membranes with the temperature close to a critical demixing point.
\\

\emph{Acknowledgement}\ \ \ \
We thank A. Travesset, P.M. Goldbart, T.C. Lubensky, and A.C. Shi for helpful discussions.
This work was supported by the National Science Foundation via Grants DMR-0605889, DMR-1106014 (RLBS, JVS) and DMR-1207026 (FY).
FY also acknowledges the support of the Center for the Physics of Living Cells of the University of Illinois at Urbana-Champaign
and the support of the Branches Cost Sharing Fund of the Institute for Complex Adaptive Matters.



\begin{thebibliography}{18}
\bibitem{incurv}
More precisely, we are interested in membranes whose curvatures are different from their spontateous values, including flat membranes with nonzero spontaneous curvatures (SC). Although this paper focuses on membranes having zero SC, the results achieved here can be easily generalized (with suitable modifications) to those with nonzero SC. In particular, when applied to flat membranes with nonzero SC, our results give the size of rafts.
\end{thebibliography}
\end{document}